\documentclass[11pt,twoside]{article}

\usepackage{asp2004}
\usepackage{epsf}
\usepackage{psfig}
\usepackage{lscape}

\begin{document}
\title{Mass loss from red giants in the Magellanic Clouds}   %%% Fill in title
\author{Jacco Th.\ van Loon}   %%% Fill in author names
\affil{Astrophysics Group, School of Physical and Geographical Sciences, Keele
University, Staffordshire ST5 5BG, UK}    %%% Fill in author affiliations

\begin{abstract} %%% Abstract to run on from here.
The Magellanic Clouds provide the ideal laboratory for the study of mass loss
and molecule and dust formation in red giants as a function of luminosity and
metal abundance. I present some recent observational work.
\end{abstract}

%%% MAIN BODY OF TEXT GOES HERE. CONSULT "INSTRUCTIONS FOR AUTHORS USING
%%% LATEX2E MARKUP", SECTIONS 2.3-2.6 FOR HELP WITH EQUATIONS, FIGURES,
%%% AND TABLES.

\section{Mass-loss rates and wind kinematics}
%%% Top level section head (remove "%" symbol)
%\subsection{}   %%% Second level section head (remove "%" symbol)
%\subsubsection{}   %%% Lowest level section head (remove "%" symbol)
%\section*{}	%%% Unnumbered top level section head (remove "%" symbol)
%\subsection*{}   %%% Unnumbered second level section head (remove "%" symbol)

Mid-infrared surveys (IRAS, MSX) have resulted in the selection of hundreds of
circumstellar dust envelopes in the Magellanic Clouds. Subsequent optical and
infrared spectroscopy and photometry are used to identify the underlying
stars, most of which are Asymptotic Giant Branch (AGB) stars or red
supergiants (van Loon et al.\ 1997, 1998a; Trams et al.\ 1999b) that exhibit
strong radial pulsations (Whitelock et al.\ 2003). The mass-loss rates are
determined by modelling the spectral energy distributions, and found to depend
on luminosity and stellar effective temperature (van Loon et al.\ 1999, 2005)
as predicted by dust-driven wind theory including multiple scattering and in
accordance with the results of hydrodynamical models. That theory also
correctly predicts the wind speed as measured from circumstellar maser
emission (van Loon et al.\ 1996, 1998b, 2001b; Marshall et al.\ 2004).
Although the dust-to-gas ratio is proportional to metal abundance the total
mass-loss rate may not depend sensitively on the initial metal abundance (van
Loon 2000). The dependence of the mass-loss rate on initial mass and
metal abundance can be calibrated by observation of mass-losing red giants
that are members of star clusters within the Magellanic Clouds (van Loon et
al.\ 2001a, 2003; van Loon et al.\ 2005a, in preparation).

%\begin{figure}[]
%\plotone{vanloonj_fig2.ps}
%\caption{Mass-losing red giants in star clusters in the Magellanic Clouds:
%(a) their initial mass and metallicity, (b) luminosity and (c) mass-loss rate.
%The solid line in panel (b) is the predicted most luminous cluster star.}
%\end{figure}

\section{Dust mineralogy and molecule formation}

Surprisingly, both oxygen-rich AGB stars and red supergiants and carbon stars
are found in the Magellanic Clouds that have very cool atmospheres despite
their low initial metal abundance (van Loon et al.\ 1998a, 2005). In
particular, it was noted that the carbonaceous molecular absorption bands in
the 3\,$\mu$m spectra of magellanic carbon stars are at least as strong as in
comparable galactic carbon stars (van Loon, Zijlstra \& Groenewegen 1999;
Matsuura et al.\ 2002, 2005). This may be due to the fact that carbon stars
produce most of the carbon themselves. Massive oxygen-rich AGB stars do not
become carbon stars due to Hot Bottom Burning (van Loon et al.\ 2001a:
$M>$4\,M$_\odot$; cf.\ Trams et al.\ 1999a), but they produce oxygen that may
also reach the stellar surface and sustain the formation of copuous amounts of
molecules. However, apparently the large molecular column density is not
reflected in an equally high efficiency of dust grain condensation (see
above), and there appears to be no strong connection between the strength of
the dust continuum emission and the strength of the molecular absorption bands
(van Loon et al.\ 2005b, in preparation). The dust detected in AGB stars and
red supergiants in the LMC is similar to that in galactic sources (Trams et
al.\ 1999b), with amorphous silicates dominating the mid-infrared spectrum of
oxygen-rich sources and silicon-carbide seen in carbon stars, but detailed
spectroscopic observations with the Spitzer Space Telescope (e.g., programme
\#3505 --- PI: Peter Wood) are expected to reveal more subtle differences
especially beyond 14\,$\mu$m.

\begin{figure}[!ht]
\plotone{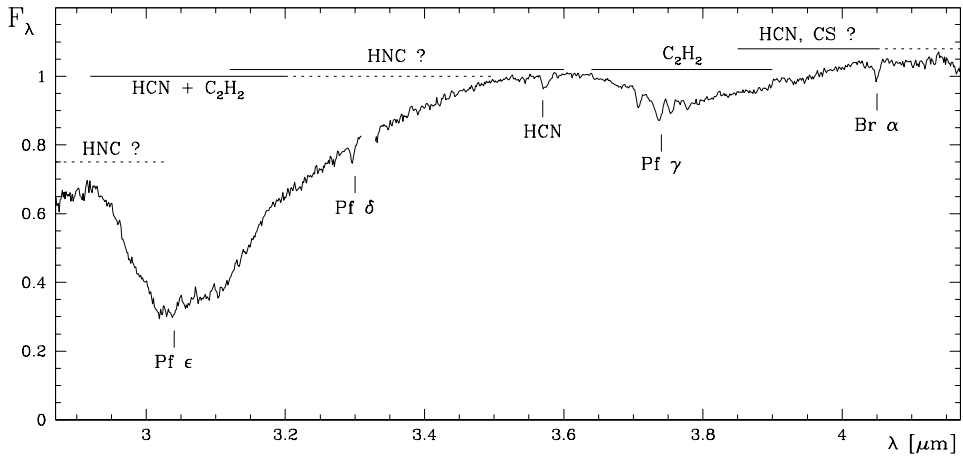}
\caption{Average 2.9--4.1\,$\mu$m VLT spectrum, for a sample of heavily
dust-enshrouded carbon stars in the LMC (van Loon et al.\ 2005b, in
preparation). Where the optical spectrum is dominated by diatomic molecules
such as C$_2$ and CN, here we see more complex molecules including C$_2$H$_2$
and HCN.}
\end{figure}

%\acknowledgements %%% Text of acknowledgements runs on after this command.

%%% THE BIBLIOGRAPHY
%%%
%%% CONSULT SECTION 3 OF "INSTRUCTIONS FOR AUTHORS" FOR HOW TO USE NATBIB.
%%% AUTHORS ARE ENCOURAGED TO USE EITHER THE "THEBIBLIOGRAPY" ENVIRONMENT
%%% BY UNCOMMENTING (DELETING THE "%" SYMBOL) THE COMMANDS BELOW, OR BY
%%% USING THE BIBTEX ENVIRONMENT. TO FIND OUT WHICH IS APPLICABLE TO YOUR
%%% CONTRIBUTION, CONSULT THE VOLUME EDITORS FOR YOUR PROCEEDINGS.
%%%


\begin{thebibliography}{}
\bibitem[]{}
Marshall J.R., van Loon J.Th., Matsuura M., et al.\ 2004, MNRAS 355, 1348
\bibitem[]{}
Matsuura M., Zijlstra A.A., van Loon J.Th., et al.\ 2005, A\&A in press
\bibitem[]{}
Matsuura M., Zijlstra A.A., van Loon J.Th., et al.\ 2002, ApJ 580, L133
\bibitem[]{}
Trams N.R., van Loon J.Th., Zijlstra A.A., et al.\ 1999a, A\&A 344, L17
\bibitem[]{}
Trams N.R., van Loon J.Th., Waters L.B.F.M., et al.\ 1999b, A\&A 346, 843
\bibitem[]{}
van Loon J.Th.\ 2000, A\&A 354, 125
\bibitem[]{}
van Loon J.Th., Zijlstra A.A., Groenewegen M.A.T.\ 1999, A\&A 346, 805
\bibitem[]{}
van Loon J.Th., Zijlstra A.A., Bujarrabal V., Nyman L.-\AA\ 1996, A\&A 306,
L29
\bibitem[]{}
van Loon J.Th., Zijlstra A.A., Whitelock P.A., et al.\ 1997, A\&A 325, 585
\bibitem[]{}
van Loon J.Th., Zijlstra A.A., Whitelock P.A., et al.\ 1998a, A\&A 329, 169
\bibitem[]{}
van Loon J.Th., te Lintel Hekkert P., Bujarrabal V., et al.\ 1998b, A\&A 337,
141
\bibitem[]{}
van Loon J.Th., Groenewegen M.A.T., de Koter A., et al.\ 1999, A\&A 351, 559
\bibitem[]{}
van Loon J.Th., Zijlstra A.A., Kaper L., et al.\ 2001a, A\&A 368, 239
\bibitem[]{}
van Loon J.Th., Zijlstra A.A., Bujarrabal V., Nyman L.-\AA\ 2001b, A\&A 368,
950
\bibitem[]{}
van Loon J.Th., Marshall J.R., Matsuura M., Zijlstra A.A.\ 2003, MNRAS 341,
1205
\bibitem[]{}
van Loon J.Th., Cioni M.-R.L., Zijlstra A.A., Loup C.\ 2005, A\&A submitted
\bibitem[]{}
Whitelock P.A., Feast M.W., van Loon J.Th., Zijlstra A.A.\ 2003, MNRAS 342, 86
\end{thebibliography}
\end{document}